\documentclass[conference,english]{IEEEtran}
\usepackage[T1]{fontenc}
\usepackage[latin9]{inputenc}
\usepackage{amsmath}
\usepackage{todonotes}
\usepackage{fancyhdr}
\usepackage[affil-it]{authblk}
\usepackage[margin=2.5cm]{geometry}
\usepackage{tabularx,caption}
\usepackage[ruled]{algorithm2e}
\usepackage{babel}
\usepackage{hyperref}
\usepackage{mathtools}

\usepackage{subfig}
\usepackage{wrapfig}
\usepackage{pbox} 
\usepackage{setspace}
\usepackage{threeparttable}
\usepackage{amssymb}
\usepackage{pifont}
\usepackage{colortbl}

\newcount\shortyear\newcount\shorthour\newcount\shortminute
\shorthour=\time\divide\shorthour by 60\shortyear=\shorthour
\multiply\shortyear by 60\shortminute=\time\advance\shortminute by
-\shortyear
\shortyear=\year\advance\shortyear by -1900

\def\zeit{\number\shorthour:\ifnum\shortminute<10 0\number\shortminute
\else\number\shortminute\fi}

\def\rightmark{{\small \today, \zeit{}}}
\newcommand{\ie}{i.\,e.\xspace}
\newcommand{\eg}{e.\,g.\xspace}

\begin{document}

\author{Gerd Lindner, Christian L. Staudt, Michael Hamann, Henning Meyerhenke, Dorothea Wagner  \smallskip\\

gerd.lindner@student.kit.edu \\  \{christian.staudt, michael.hamann, meyerhenke, dorothea.wagner\}@kit.edu \\
Institute of Theoretical Informatics, Karlsruhe Institute of Technology (KIT), Germany \\
Am Fasanengarten 5, 76131 Karlsruhe -- Tel: +49 721 60841821
}

\title{Structure-Preserving Sparsification \\ of Social Networks}

\date{}

\maketitle

\begin{abstract}
Sparsification reduces the size of networks while preserving structural and statistical properties of interest.
Various sparsifying algorithms have been proposed in different contexts.
We contribute the first systematic conceptual and experimental comparison of \textit{edge sparsification} methods on a diverse set of network properties.
It is shown that they can be understood as methods for rating edges by importance and then filtering globally by these scores.
In addition, we propose a new sparsification method (\textit{Local Degree}) which preserves edges leading to local hub nodes.
All methods are evaluated on a set of 100 Facebook social networks with respect to network properties including diameter, connected components, community structure, and multiple node centrality measures.
Experiments with our implementations of the sparsification methods (using the open-source network analysis tool suite NetworKit) 
show that many network properties can be preserved down to about 20\% of the original set of edges. 
Furthermore, the experimental results allow us to differentiate the behavior of different methods and show which method is suitable with respect to which property.
Our Local Degree method is fast enough for large-scale networks and performs well across a wider range of properties than previously proposed methods.

\end{abstract}

\section{Introduction}

\subsection{Context}

Complex networks have nontrivial structures and statistical properties and are often represented by graphs.
Such data models have been employed in countless domains based on the observation that the structure of relationships yields insights into the composition and behavior of complex systems~\cite{costa2011analyzing}.
Many concepts were pioneered in the study of social networks, in which edges represent social ties between social actors.
Most real-world complex networks, including social networks, are already sparse in the sense that for $n$ nodes the edge count $m$ is asymptotically in $O(n)$.
Nonetheless, typical densities lead to a computationally challenging number of edges.
Here we pursue the goal of further sparsifying such networks by retaining just a fraction of edges (sometimes called a ``backbone'' of the network), while showing experimentally that important properties of networks can be preserved in the process.

Potential applications of network sparsification are numerous.
One of them is information visualization: 
Even moderately sized networks turn into ``hairballs'' when drawn with standard techniques, as the amount of edges is visually overwhelming.
In contrast, showing only a fraction of edges can reveal network structures to the human eye if these edges are selected appropriately. 
Sparsification can also be applied as an acceleration technique: By disregarding a large fraction of edges that are unimportant for the task, running times of graph and network analysis algorithms can be reduced.
From a network science perspective, sparsification can yield valuable insights into the importance of relationships and the participating nodes: 
Given that a sparsification method tends to preserve a certain property, the method can be used to rank or classify edges, discriminating between essential and redundant edges.
Many other possible applications arise if we think of sparsification as lossy compression.
Large networks can be strongly reduced in size if we are only interested in certain structural aspects that are preserved by the sparsification method.

The core idea of the research presented here is that not all edges are equally important with respect to properties of a network: For example, a relatively small fraction of long-range edges typically act as shortcuts and are responsible for the small-world phenomenon in complex networks. 
The importance of edges can be quantified, leading to \textit{edge scores}, often also referred to as \textit{edge centrality values}.
In general, we subsume under these terms any measure that quantifies the importance of an edge depending on its position within the network structure.
Sparsification can then be broken down into the stages of (i) edge scoring and (ii) filtering the edges using a global score threshold.

Despite the similar terminology, our work is only weakly related to a line of research in theoretical computer science where \textit{graph sparsification} is understood as the reduction of a dense graph ($\Theta(n^2)$ edges) to a sparse graph ($O(n)$ 
edges) while provably preserving properties such as spectral properties (\eg~\cite{batson2013spectral}).
The networks of our interest are already sparse in this sense.
With the goal of reducing network data size while keeping important properties, our research is related to a body of work that considers sampling from networks (on which~\cite{ahmed2014network} provides an extensive overview).
Sampling is concerned with the design of algorithms that select edges and/or nodes from a network.
Here, node and edge sampling methods must be distinguished:
For node sampling, nodes \emph{and} edges from the original network are discarded, while edge sampling preserves all nodes and reduces the number of edges only.
The literature on node sampling is extensive, while pure edge sampling and filtering techniques have not been considered as often.
A seminal paper~\cite{Leskovec2006} concludes that node sampling techniques are preferable, but considers few edge sampling techniques.
The study presented in~\cite{ebbes2008sampling} looks at how well a sample of 5\%-20\% of the original network preserves certain properties, and is mainly focused on node sampling through graph exploration. It concludes that random walk-based node sampling works best on complex networks, but does so on the basis of experiments on synthetic graphs only and compares only with very simple edge sampling methods. 

Only edge sampling techniques are directly comparable to our edge scoring and filtering methods.
In this work, we restrict ourselves to reducing the edge set, while keeping all nodes of the original graph.
Preserving the nodes allows us to infer properties of each node of the original graph.
This is important because in network analysis, the unit of analysis is often the individual node, \eg when a score for each user in an online social network scenario shall be computed.
With respect to the goal of accelerating the analysis, many relevant graph algorithms scale with $m$ rather than $n$, so reducing $m$ is more relevant.

Another related approach is the \textit{Multiscale Backbone}~\cite{Serrano09}, which is applicable on weighted graphs only and is therefore not included in our study.
Instead of applying a global edge weight cutoff for edge filtering, which hides important structures at different scales, this approach aims at preserving them at all scales.

\subsection{Contribution}

We contribute the first systematic conceptual and experimental comparison of existing and novel edge scoring and filtering methods on a diverse set of network properties.
Descriptions and literature references for the related methods which we reimplemented are given in Section~\ref{sec:methods}.
Our results illuminate which methods are suitable with respect to which properties of a network.
In particular, the Local Degree method we propose is based on simple principles but surprisingly effective across a wider range of properties than previously proposed methods.
Furthermore, upon acceptance, we publish efficient parallelized implementations and a framework for such methods as part of 
the \textit{\href{http://networkit.iti.kit.edu}{NetworKit}} open-source tool suite~\cite{Staudt14}.
While our study covers various approaches from the literature, it is by no means exhaustive due to the vast amount of potential sparsification techniques. 
With future methods in mind, we hope to contribute a framework for their implementation and evaluation.

%
\section{Network Properties}
\label{sec:prop}

The structure of a complex network is usually characterized in terms of certain key figures and statistics~\cite{newman2010networks}.
Decomposition of the network into cohesive regions is a frequent analysis task: 
All nodes in a \textit{connected component} are reachable from each other.
\textit{Communities} are subsets of nodes that are internally dense and externally sparsely connected.
The \textit{diameter} of a graph is the length of its longest shortest path.
The observation that the diameter of social networks is often surprisingly small is referred to as the \textit{small world phenomenon}.
In case of disconnected graphs, we consider the diameter of the largest component.

Node centrality measures quantify the relative importance of a node within the network structure. 
The distribution of \textit{degrees}, the number of connections per node,  plays  an important role in characterizing a network: Empirically observed complex networks tend to show a heavy 
tailed \textit{degree distribution} which follows a power-law with a characteristic exponent: $p(k) \sim k^{-\gamma}$.  
\textit{Clustering coefficients} are key figures for the amount of transitivity in networks, \ie the tendency of edges to form between indirect  neighbor nodes. 
\textit{Betweenness centrality} expresses the concept that a node is important if it lies on many shortest paths between nodes in the network.
\textit{PageRank} assigns relative importance to nodes according to their connections, incorporating the idea that edges to high-scoring nodes contribute more.
While this collection is not and cannot be exhaustive, we choose these common measures for our experimental study (Section~\ref{sec:exp}).

\section{Sparsification Methods}
\label{sec:methods}

All sparsification methods we consider can be split up into two stages:
(i) the calculation of a score for each of the $m$ edges in the input graph (where the score is high if the edge is important) and (ii) subsequent global filtering according to these scores.
In this section we present the existing and new approaches we consider and show for each of these methods how it can be transformed into an edge score that can be used for global filtering.

\paragraph*{Random Edge (RE)}

When studying different sparsification algorithms, the performance of random edge selection is an important baseline. 
As we shall see, it also performs surprisingly well.
The method selects edges uniformly at random from the original set such that the desired sparsification ratio is obtained. 
This is equivalent to scoring edges with values chosen uniformly at random.
Naturally this needs time linear in the number of edges.

\paragraph*{Triangles}
Especially in social networks, triangles play an important role because the presence of a triangle indicates a certain quality of the relationship between the three involved nodes.
The sociological theory of Simmel~\cite{simmel1950sociology} states that ``triads (sets of three actors) are fundamentally different from dyads (sets of two actors) by way of introducing mediating effects.''
In a friendship network, it is likely for two actors with a high number of common friends to be friends as well.
Filtering globally by triangle counts tends to destroy local structures, but several of the following sparsification methods are based on the triangles edge score $T(u, v)$ that denotes for an edge $\{u, v\}$ the number of triangles it belongs to.
The time needed for counting the number of all triangles is $O(m\cdot a)$~\cite{ortmann2014triangle}, where $a$ is the graph's arboricity~\cite{Chi85}.

\paragraph*{Local Similarity (LS)}
\label{par:LocalSim}

One line of research attempts to sparsify graphs with the goal of speeding up data mining algorithms.
Satuluri et al.\ \cite{Satuluri2011} propose a local graph sparsification method with the intention of speedup and quality improvement of community detection.
They suggest reducing the edge set to 10-20\% of the original graph and use the Jaccard measure to quantify the overlap between node neighborhoods $N(u)$, $N(v)$ and thereby the similarity of two given nodes:

\begin{equation*}
	\operatorname{J}(u,v) = \frac{|N(u) \cap N(v)|}{|N(u) \cup N(v)|} = \frac{T(u,v)}{d(u) + d(v) - T(u, v)},
\end{equation*}

\noindent where $d(u)$ denotes the degree of $u$.
Global sparsification approaches tend to destroy small network structures that are relevant from a local point of view.
In order to achieve local instead of global sparsification, Satuluri et al.\ keep for each node $u$ the top $\lfloor d(u)^\alpha \rfloor$ edges incident to $u$, ranked according to their similarity ($\alpha \in [0,1]$).
Note that this procedure ensures that at least one incident edge of each node is retained.
This is equivalent to assigning each edge the score $1- \alpha$ for the minimum value of $\alpha$ such that the edge is kept in the sparsified graph and filtering by this edge score.
The time needed for calculating this edge score is the time for counting all triangles and for sorting the neighbors of all nodes, which can be done in $O(m \log(d_{\max}))$. The authors also propose a fast approximation which runs in time $O(m)$.
This sparsification technique has also been adapted for accelerating \textit{collective classification}, \ie the task of inferring the labels of all nodes in a graph given a subset of labeled nodes~\cite{saha2013sparsification}.

\paragraph*{Simmelian Backbones (TS, QLS)}

The \textit{Simmelian Backbones} introduced by Nick et al.~\cite{Nick13} aim at discriminating between edges that are placed within dense subgraphs and those between them.
The original goal of these methods was to produce readable layouts of networks.
To achieve a ``local assessment of the level of actor neighborhoods''~\cite{Nick13}, 
the authors propose the following approach, which we adapt to our concept of edge scores. Given an edge scoring method $S$ and a node $u$, they introduce the notion of a rank-ordered neighborhood as the list of adjacent neighbors sorted by $S(u,\cdot)$ in descending order.
The original \textit{(Triadic) Simmelian Backbone} uses triangle counts $T$ for $S$.
The newer \textit{Quadrilateral Simmelian Backbone} by Nocaj et al.~\cite{nocaj2014hairballs} uses \textit{quadrilateral edge embeddedness}, which they define as
\begin{equation*}
Q(u,v) = \frac{q(u,v)}{\sqrt{q(u) \cdot q(v)}}
\end{equation*}
with $q(u,v)$ being the number of quadrangles containing edge $\{u,v\}$ and $q(u)$ being the sum of $q(u,v)$ over all neighbors $v$ of $u$.
They argue that this modified version performs even better at discriminating edges within and between dense subgraphs.

On top of the rank-ordered neighborhood graph that is induced by the ranked neighborhoods of all nodes, Nick et al.\ introduce two filtering techniques, a parametric one and a non-parametric one. Like Nocaj et al.\ we use only the non-parametric variant. By \textit{TS}, we denote the Triadic Simmelian Backbone and by \textit{QLS} the Quadrilateral Simmelian Backbone.
The non-parametric variant uses the Jaccard measure similar to Local Similarity but, instead of considering the whole neighborhood, they use the maximum of the Jaccard measure of the top-$k$ neighborhoods for all possible values of $k$.
While the time needed for quadrangle counting is equal to the time for triangle counting~\cite{Chi85}, the overlap and Jaccard measure calculation of prefixes needs time $O(m \cdot d_{\max} \log(d_{\max}))$ as it needs to be separately calculated for all edges.

\paragraph*{Edge Forest Fire (EFF)}

The original Forest Fire node sampling algorithm~\cite{Leskovec2006} is based on the idea that nodes are ``burned'' during a fire that starts at a random node and may spread to the neighbors of a burning node. 
Note that contrary to random walks the fire can spread to more than one neighbor but already burned neighbors cannot be burned again. The basic intuition is that nodes and edges that get visited more frequently than others during these walks are more important.
In order to filter edges instead of nodes, we introduce a variant of the algorithm in which we use the frequency of visits of each edge as a proxy for its relevance.
As the total length of all walks is hard to estimate in advance,
we cannot give a tight bound for the running time.

\begin{figure}[h!]
\includegraphics[width=0.21\textwidth]{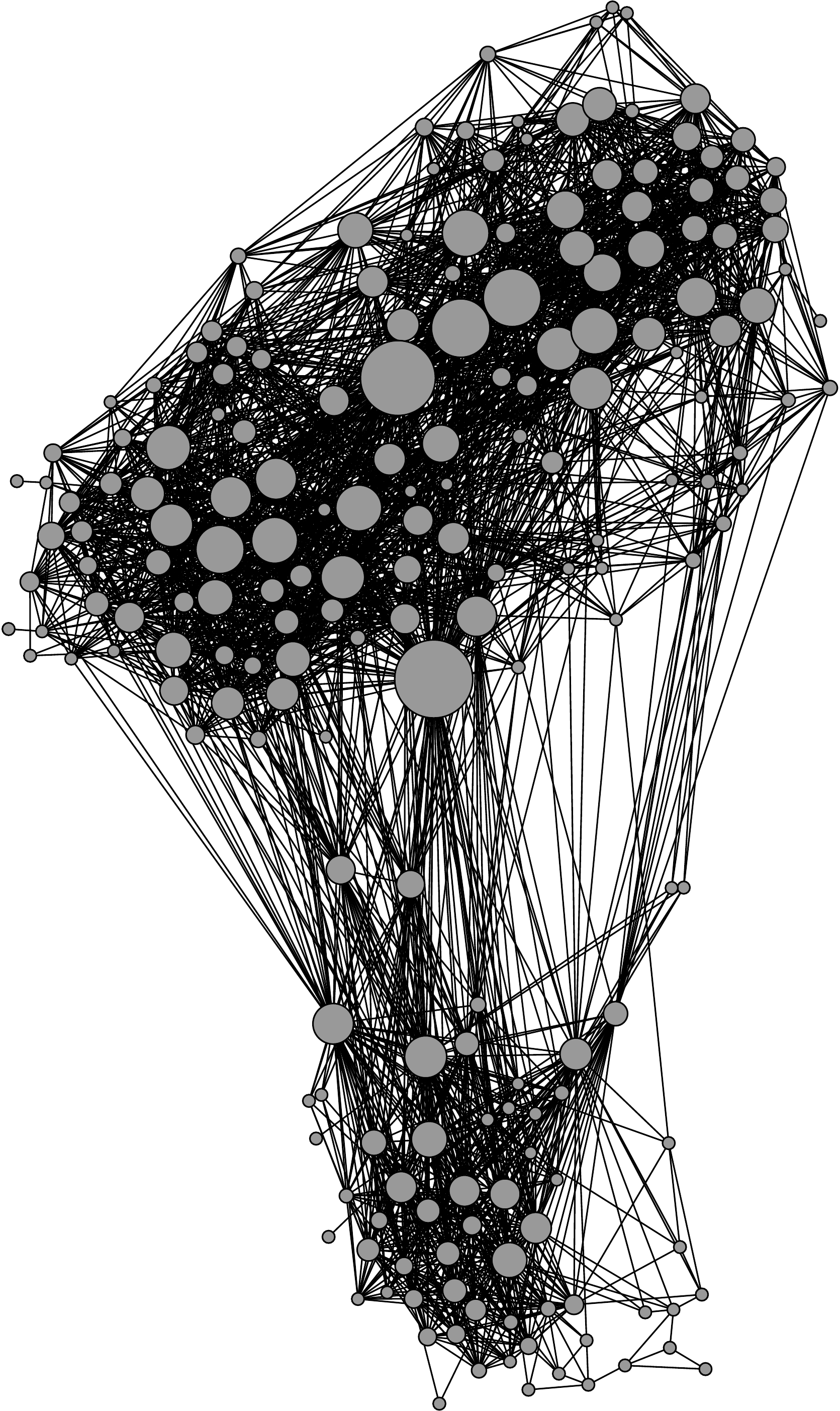}
\includegraphics[width=0.21\textwidth]{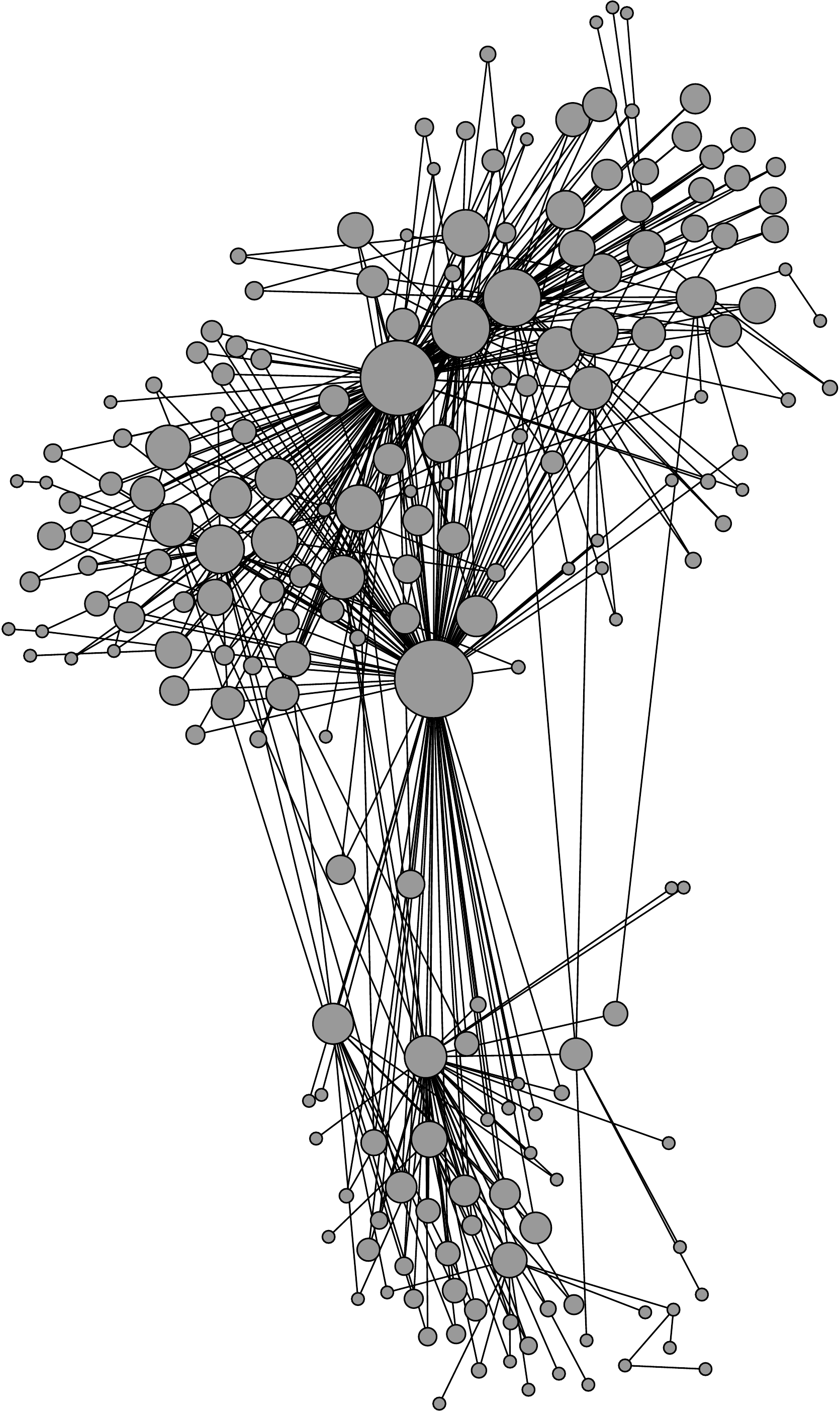}
\caption{Drawing of the Jazz musicians collaboration network and the \textit{Local Degree} sparsified version containing 15\% of edges. Node size proportional to degree.}
\label{fig:jazz}
\end{figure}

\paragraph*{Local Degree (LD)}
Inspired by the notion of hub nodes, \ie nodes with locally relatively high degree, as well as the approach of Satuluri et al.~\cite{Satuluri2011} 
with their Local Similarity method, we propose the following new sparsification method:
For each node $v \in V$, we include the edges to the top $\lfloor \deg(v)^\alpha \rfloor$ neighbors, sorted by degree in descending order.
Similar to Local Similarity we use again $1-\alpha$ for the minimum parameter $\alpha$ such that an edge is still contained in the sparsified graph as edge score.
The goal of this approach is to keep those edges in the sparsified graph that lead to nodes with high degree, i.e.\ the hubs that are crucial for a complex network's topology.
The edges left after filtering form what can be considered a ``hub backbone'' of the network (see Fig.~\ref{fig:jazz} for an example).

As only the neighbors of each node need to be sorted, this can be done in $O(m\log(d_{\max}))$. 
Using linear-time sorting it is even possible in $O(m)$ time.

\section{Implementation}
\label{sec:impl}

For this study, we created efficient C++ implementations of all considered sparsification methods, 
and accelerated them using OpenMP parallelization.
In particular, RE, LD, LS and the Simmelian Backbone methods (with exception of the inherently sequential triangle and quadrangle counting algorithms~\cite{Chi85}) have been parallelized.
We implemented the algorithms in \emph{NetworKit}~\cite{Staudt14}, an interactive tool suite for scalable network analysis. It provides a large set of graph algorithm implementations we used for our experiments. NetworKit combines kernels written in C++ with an interactive Python shell to achieve both high performance and interactivity, a concept we use for our implementations as well. 
For community detection, we use an efficient implementation of the Louvain method that is also part of NetworKit~\cite{staudt2015engineering}.
To get consistent results, a deterministic configuration of this algorithm is used.

\textit{Gephi}~\cite{BastianHJ09Gephi} is a graph visualization tool which we use not only for visualization purposes but also for interactive exploration of sparsified graphs. To achieve said interactivity, we implemented a client for the \emph{Gephi Streaming Plugin} in NetworKit. It is designed to stream graph objects from and to Gephi utilizing the JSON format. Using our implementation in NetworKit, a few lines of Python code suffice to sparsify a graph, calculate various network properties, and export it to Gephi for drawing.
The approach of separating sparsification into edge score calculation and filtering allows for a high level of interactivity by exporting edge scores from NetworKit to Gephi and dynamic filtering within Gephi. 

%
\section{Experimental Study}
\label{sec:exp}
\subsection{Quantifying Similarity in Network Properties}
\label{sec:quantifying}

Quantifying the similarity between a network and its sparsified version is an intricate problem. 
Ideally, a similarity measure should meet the following requirements:
a) \emph{Ignoring trivial differences}: Consider, for example, the degree distribution: One cannot expect the distribution to remain identical after edges get removed during sparsification.
It is clear, however, that the general shape of the distribution should remain ``similar'' and that high-degree nodes should remain high-degree nodes in order to consider the degrees as preserved.
b) \emph{Intuitive and Normalized}: Similarity values from a closed domain like $[0,1]$ allow for aggregation and comparability.
A similarity value of $1$ indicates that the property under consideration is fully preserved, whereas a value of $0$ indicates that similarity is entirely lost.
c) \emph{Revealing Method Behavior}: A good similarity measure will clearly expose different behavior between sparsification methods.
d) \emph{Efficiently computable}. 

Following these requirements, we select the following measures:
In order to observe how the network diameter changes through sparsification (Sec.~\ref{sec:results}), we plot the quotient of the original network diameter and the resulting diameter, which yields legible results since in practice the diameter does not decrease during sparsification.
Both the detection of connected components and communities yield partitions of the node set into disjoint subsets.
We use Normalized Mutual Information (NMI) as a similarity value, a common measure for comparing partitions of graphs~\cite{Lancichineti2008}.
Node degree, betweenness and PageRank can be treated as node centrality indices which represent a ranking of nodes by structural importance.
Since absolute values of the centrality scores are less interesting than the resulting rank order,
we compare the rankings before and after sparsification using Spearman's $\rho$ rank correlation coefficient.
(This focus on rank order is also the reason why we did not adopt the Kolmogorov-Smirnov statistic used in~\cite{Leskovec2006}, which compares distributions of absolute values.)
Even though the local clustering coefficient can be interpreted as a centrality score as well, the comparison of ranks does not seem meaningful in this case due to the fact that it is a local score. Instead, we analyse the deviation of the average local  clustering coefficient from the original value. 

\subsection{Setup}
Experiments were performed on a multicore compute server with 4 physical Intel Core i7 cores at 3.4 GHz, 8 threads, and 32 GB of memory.
For this explorative study, we use a collection of 100 social networks representing early snapshots of Facebook, each of which is a student online friendship network for a US university~\cite{traud2012social}.
Sizes of the networks are between 10k and 1.6 million edges.
For the plots in Sec.~\ref{sec:results}, we aggregate experimental results over this set.
We chose to focus on a set of networks of one type, i.e.\ with a common origin and high structural similarity among the networks, in order to get meaningful aggregated values.
It remains an open question to what extent results can be translated to other types of complex networks, since according to experience the performance of network analysis algorithms depends strongly on the network structure.

\subsection{Correlations between Edge Scores}
\label{sub:correlation}

\begin{figure}[h!]
\begin{center}
\includegraphics[width=\columnwidth]{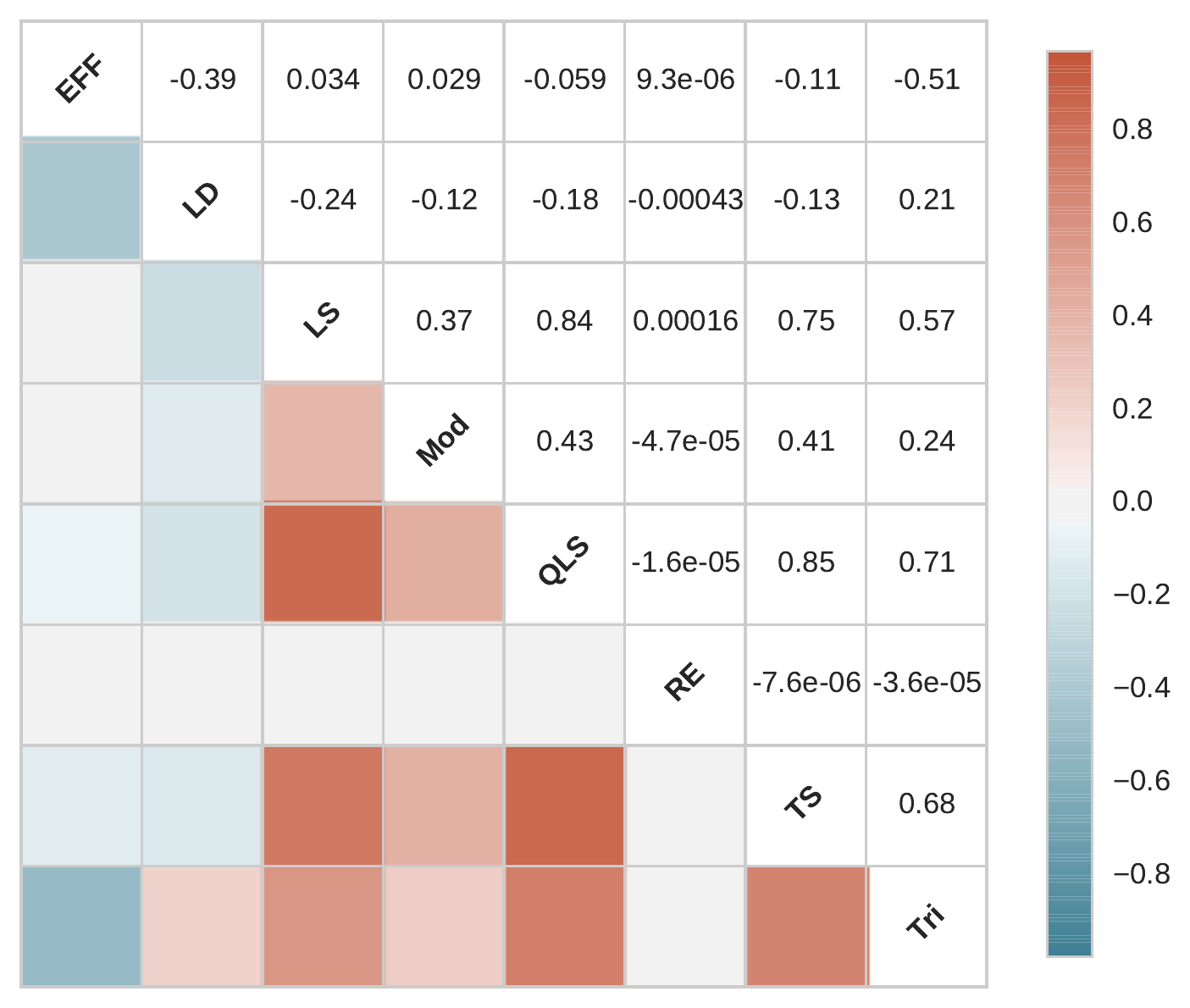}
\caption{Edge score correlations (Spearman's $\rho$)}
\label{fig:corrFb}
\end{center}
\end{figure}

Among our sparsification methods, some are more similar to others in the sense that they tend to preserve similar edges. 
Such similarities can be clarified by studying correlations between edge scores.
We calculate edge score correlations for the set of 100 social networks as follows:
For each single network, edge scores are calculated with the various scoring methods and Spearman's rank correlation coefficient is applied. 
The coefficient is then averaged over all networks and plotted in the correlation matrix (Figure~\ref{fig:corrFb}).
There is one column for each method, and the column \textsf{Mod} represents edge scores that are 1 for intra-community edges and 0 for inter-community edges after running a modularity-maximizing community detection algorithm.
Positive correlations with these scores indicate that the respective rating method assigns high scores to edges within modularity-based communities. 

Interpretation of the results is challenging: 
The correlations we observe reflect intrinsic, mathematical similarities of the rating algorithms on the one hand, but on the other hand they are also caused by the structure of this specific set of social networks (e.g., it may be a characteristic of a given network that edges leading to high-degree nodes are also embedded in many triangles).
Nonetheless, we note the following observations:
The methods intended to preserve edges within dense subgraphs (LS, QLS, TS) are clearly positively correlated, and also have
positive correlations with modularity-based communities and the number of triangles an edge is embedded in (Tri).
Interestingly, a strong positive correlation exists between Local Similarity and Quadrilateral Simmelian Backbone, which have different computational costs, but are predicted to show similar sparsifying effects.
The two new methods we introduce, Edge Forest Fire and Local Degree, set themselves apart from this class, while also being negatively correlated with each other. 
LD tends to favor edges between dense regions, unlike the LS and Simmelian methods. 
The strong negative correlation between EFF and triangle count can be explained by the fact that the Edge Forest Fire can never ``burn'' a triangle, as nodes cannot be visited twice.

\subsection{Preservation of Properties}
\label{sec:results}

In the following plots, the measures discussed in Sec.~\ref{sec:quantifying} are shown on the y-axis for a given ratio of kept edges  ($m' / m$) on the x-axis (e.g., a ratio of 0.2 means that 20\% of edges are still present).

\paragraph*{Diameter}

\begin{figure}[h!]
\begin{center}

\subfloat[Original network diameter divided by network diameter ]{
\includegraphics[width=\columnwidth]{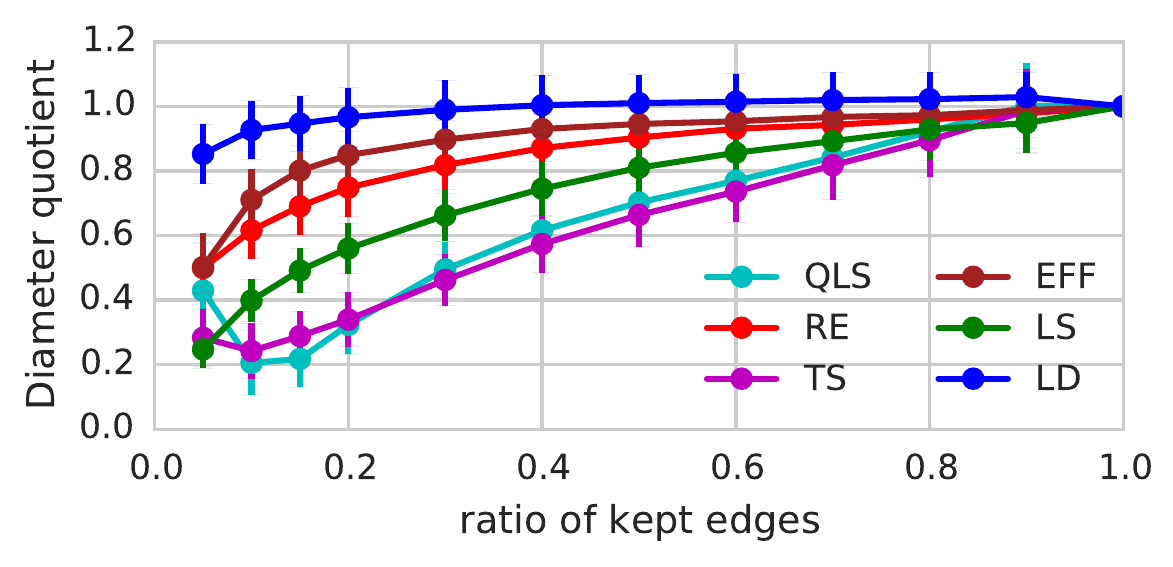}
\label{fig:plt_diameter}
}

\subfloat[Deviation from original clustering coefficient]{
\includegraphics[width=\columnwidth]{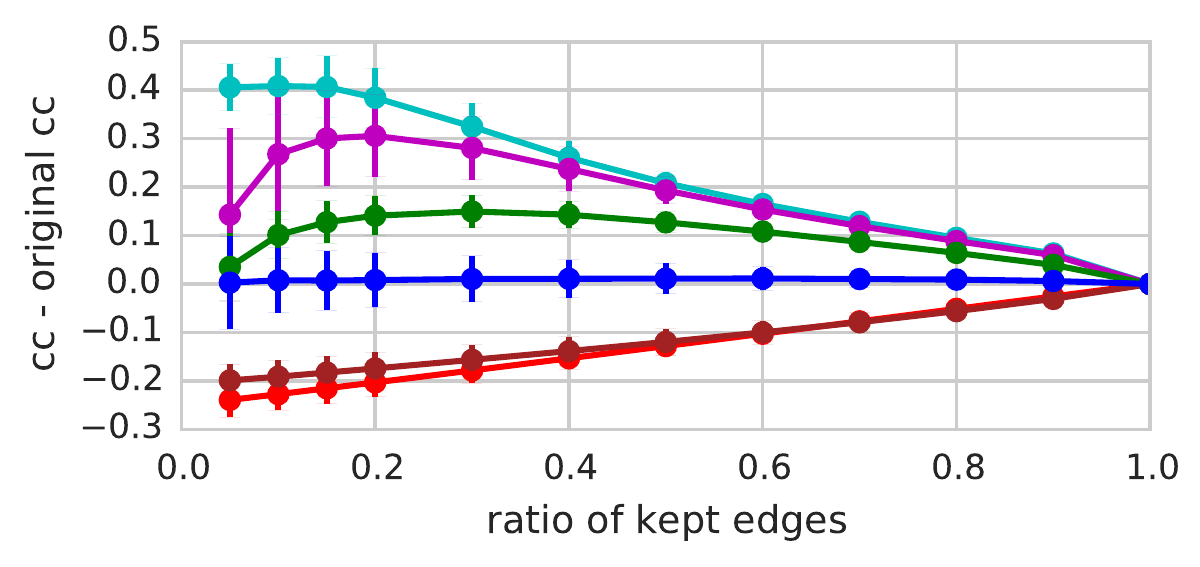}
\label{fig:plt_ccAvgLocal}
}%

\caption{Preservation of global network properties}

\end{center}
\end{figure}

We motivated the Local Degree method with the idea that shortest paths commonly run through hub nodes in social networks.
Therefore, preserving edges leading to high-degree nodes should preserve the small diameter.
This is confirmed by our experiments (Figure \ref{fig:plt_diameter}).
In contrast, methods that prefer edges within dense regions clearly do not preserve the diameter.
With Simmelian Backbones the diameter drops when only few edges are left; this can be explained by the fact that Simmelian Backbones do not maintain the connectivity and that at the end the graph is decomposed into multiple connected components which have a smaller diameter.

\paragraph*{Clustering Coefficient}
As far as the average local clustering coefficient is concerned, we observe three classes of sparsification methods. Figure \ref{fig:plt_ccAvgLocal} shows the deviation from the original value, averaged over all graphs in our dataset. For both RE and EFF, which are based on randomness, the clustering coefficient drops almost linearly with decreasing sparsification ratio. TS, QLS and LS keep mostly edges within dense regions, which results in increasing clustering coefficients. Note that our method LD keeps the deviation close to zero for sample sizes down to 10\%.

\paragraph*{Node Centrality Measures}

\begin{figure}[h!]
\begin{center}

\subfloat[Spearman's rank correlation coefficient for node degree ]{
\includegraphics[width=\columnwidth]{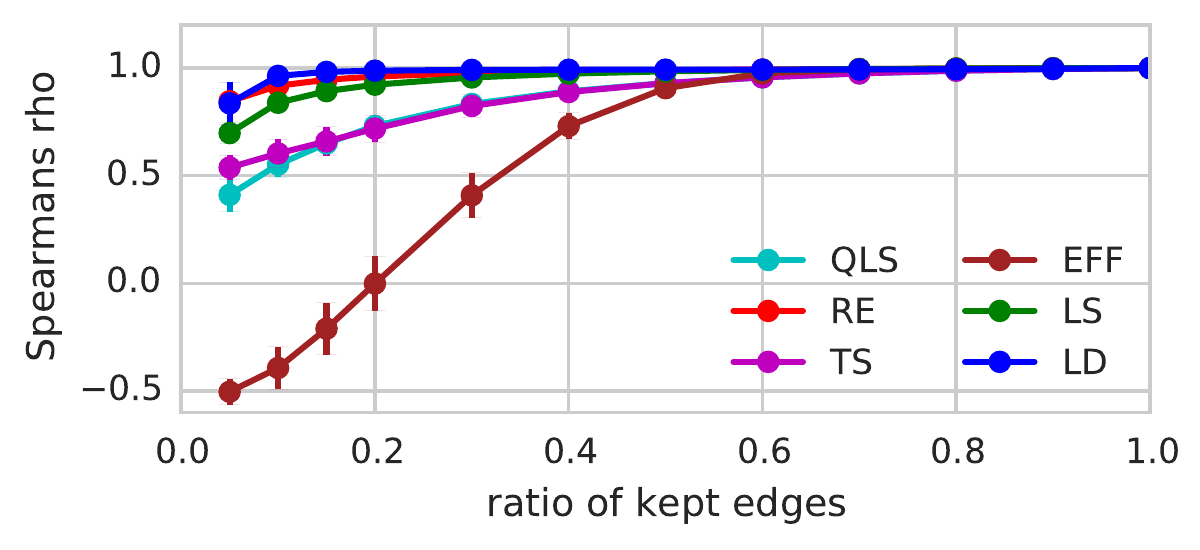} 
\label{fig:plt_dd_spearman_rho}
}%

\subfloat[Spearman's rank correlation coefficient for betweenness centrality ]{
\includegraphics[width=\columnwidth]{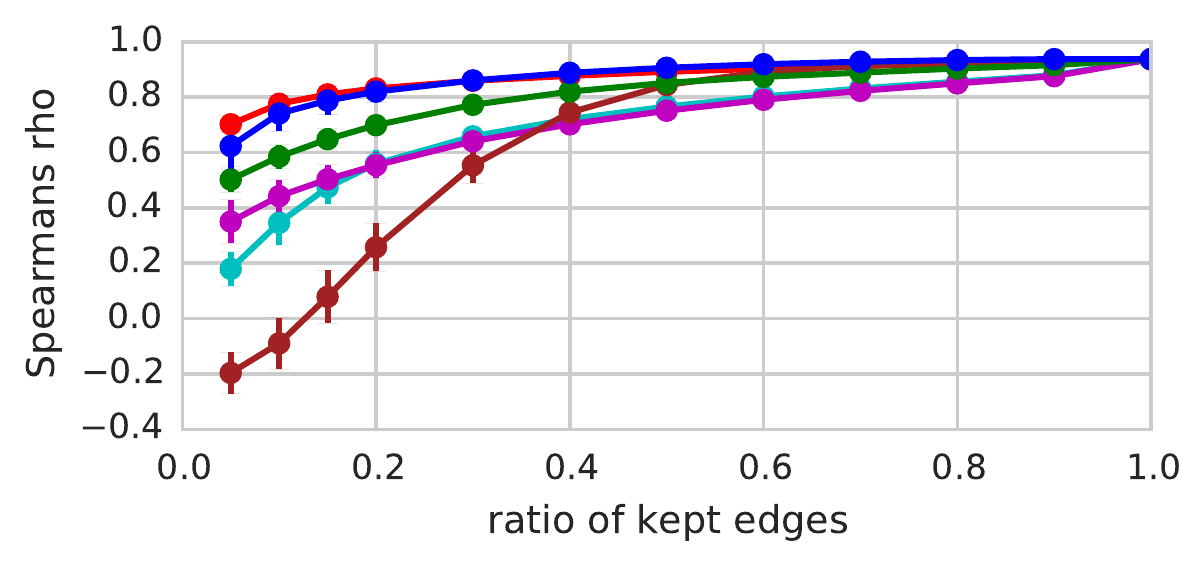}
\label{fig:plt_betweenness_spearman_rho}
}%

\subfloat[Spearman's rank correlation coefficient for PageRank centrality ]{
\includegraphics[width=\columnwidth]{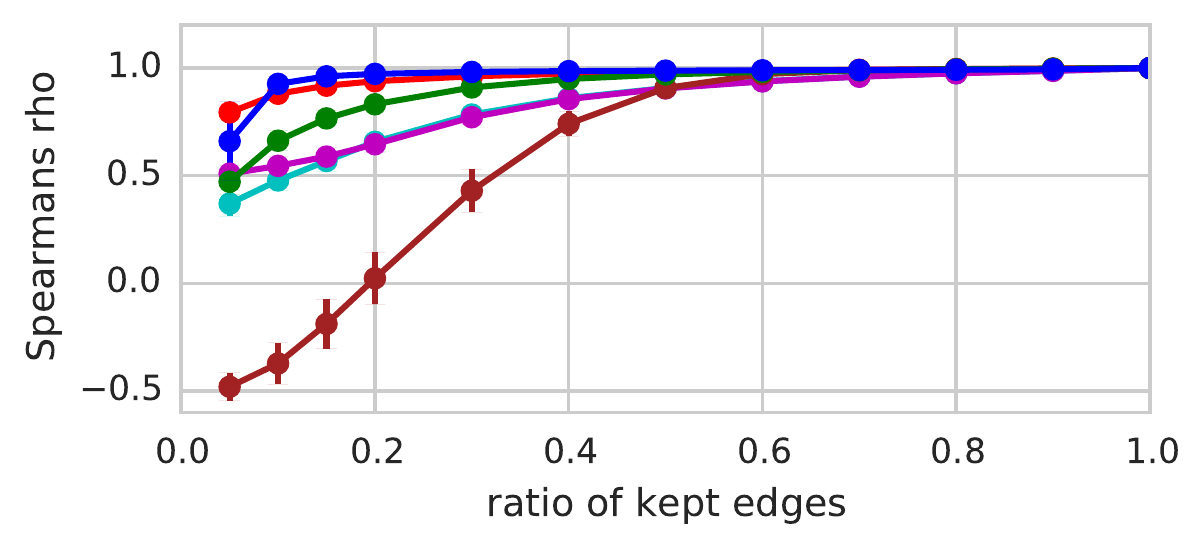}
\label{fig:plt_cc_spearman_rho}
}%

\caption{Preservation of node centrality measures}
\label{fig:centralities}

\end{center}
\end{figure}

The similarity of curves in Figure~\ref{fig:centralities} catches the eye immediately:
For these node centrality measures, the sparsification methods behave in a very similar way, with random edge deletion and Local Degree performing best and Edge Forest Fire failing early.
This similarity could be explained by strong correlations between node degree, PageRank and betweenness, which have been observed before (e.g.~\cite{fortunato2008approximating}).
Likewise, EFF fails because it cannot preserve node degrees, as the expected number of randomly selected incident edges via the ``burning process'' is relatively low even for high-degree nodes.
In accordance with our intuition that edges leading to high-degree neighbors are important and should be preserved, our experiments show that the Local Degree method preserves multiple node centralities (i.e.\ degree, PageRank and betweenness).
Random Edge filtering is also quite good at preserving betweenness, PageRank and degree centrality.
Again, methods that are focused on keeping edges within dense regions are not as good at preserving said properties.
However, filtering locally seems to help the Local Similarity sparsification technique to perform still better than Simmelian Backbones.

\paragraph*{Components and Communities}

\begin{figure}[h!]
\begin{center}

\subfloat[Normalized Mutual Information between partitions into communities ]{
\includegraphics[width=\columnwidth]{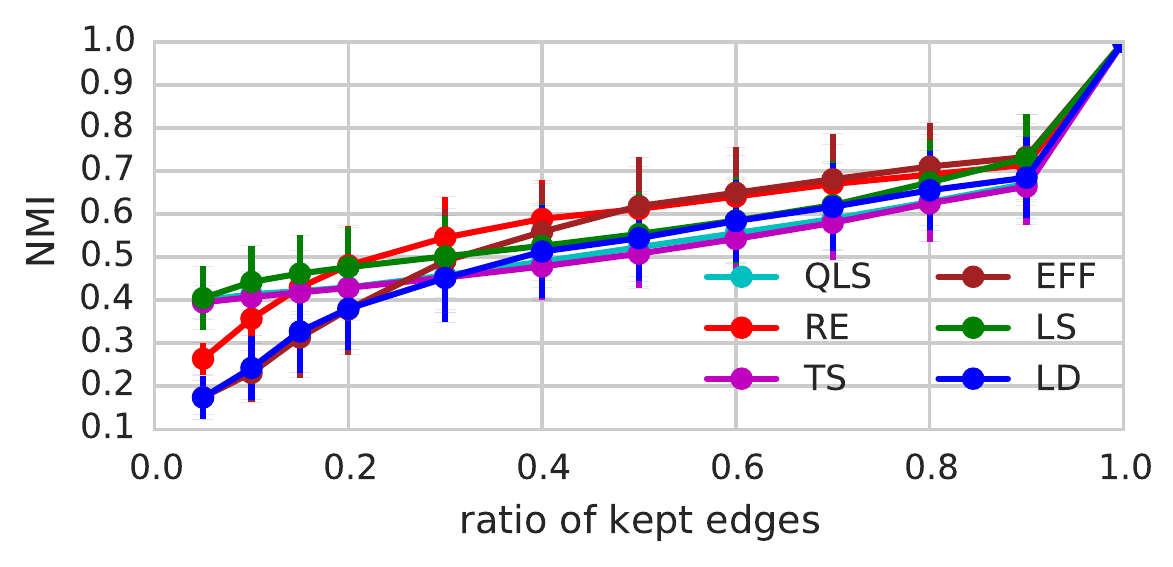}
\label{fig:plt_nmi}
}

\subfloat[Number of communities in original graph divided by number of communities in sparsified graph (according to PLM community detection, average over all graphs)]{
\includegraphics[width=\columnwidth]{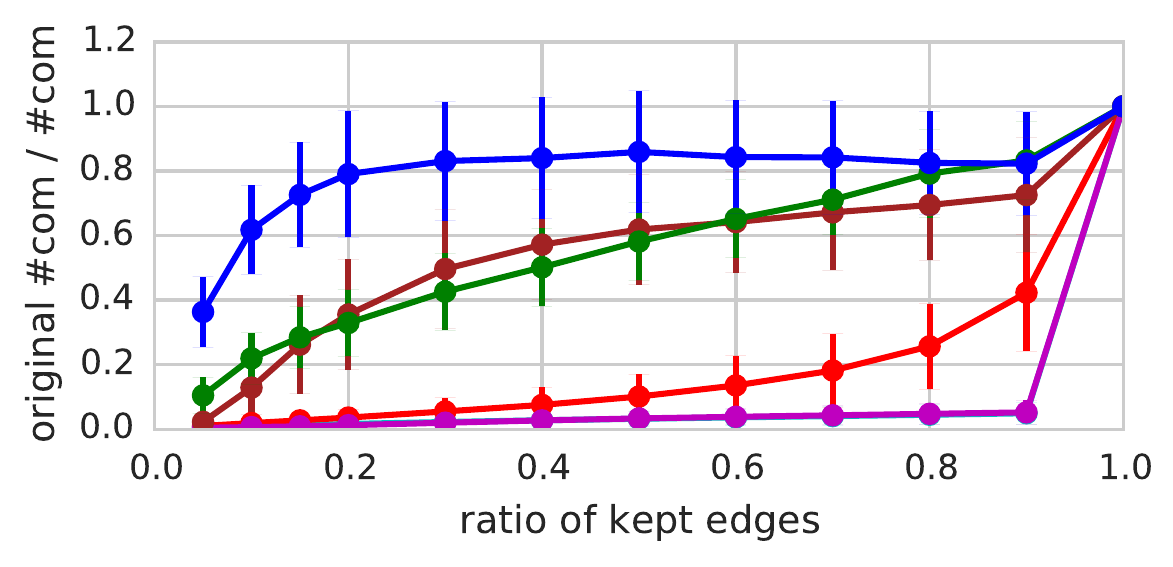}
\label{fig:plt_numCommunities}
}

\subfloat[NMI of connected components ]{
\includegraphics[width=\columnwidth]{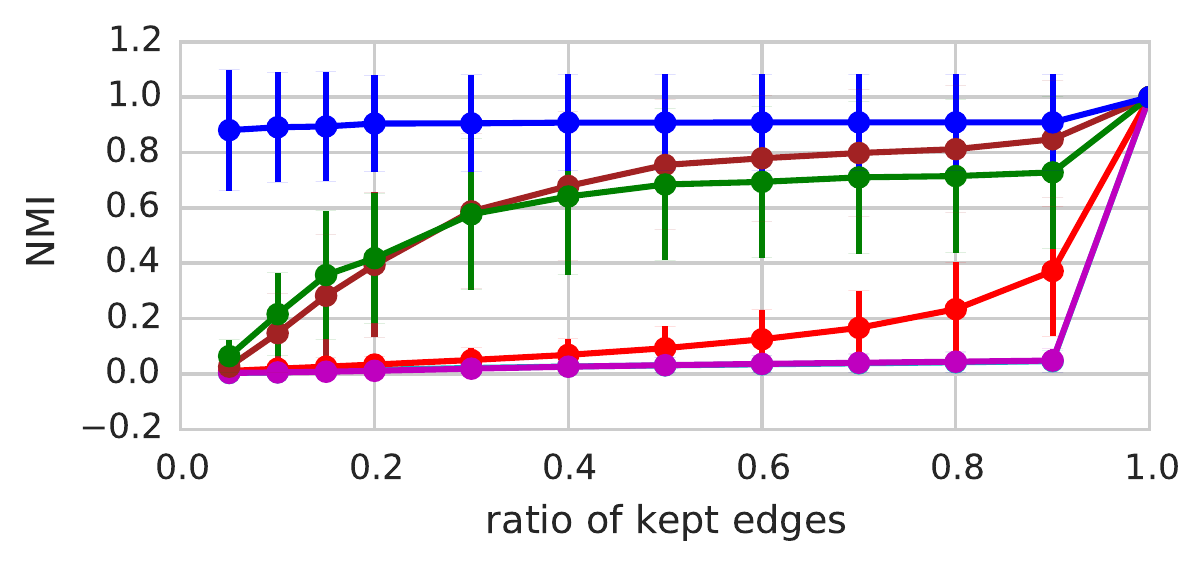}
\label{fig:plt_wcc_nmi}
}

\caption{Preservation of network cohesion}

\end{center}
\end{figure}

As shown in Figure~\ref{fig:plt_wcc_nmi}, our Local Degree method best preserves the connected components of the graph.
Random edge filtering and Simmelian Backbones on the contrary tend to separate nodes of degree 1 from the rest of the graph.
Local Similarity also preserves connectivity but as its retained edges are not directed towards central hubs, it easily disconnects small groups of nodes.
According to the NMI measure in Figure~\ref{fig:plt_nmi}, it seems that random edge sampling is best suited for preserving the community structure as it is found by the Louvain method.
However, if we consider the number of communities in Figure~\ref{fig:plt_numCommunities}, the results are quite different.
The Simmelian Backbones generate singletons rather quickly.
This explains why the number of communities increases so quickly.
Random edge filtering leads to the same phenomenon.
Nevertheless, the communities found seem to differ significantly for all sparsification methods.
The Local Degree sparsification method is the only method 
we consider that is able to keep the number of communities relatively unchanged up to a very high degree of sparsification. The nonetheless rather low NMI similarity values can be explained by the following behavior: 
Consider a hub node $x$ within a community with neighbors that are for the most part also connected to a hub node $y$ with higher degree than $x$. 
Due to the way Local Degree scores edges, $x$ will lose many of its connections within the community and may be pulled into the community of a neighboring high-degree node $z$ that is not part of the original community of $x$.  
As most real-world networks do not have one community structure but many, it has to be left as an open question if those sparsification techniques that keep the number of communities within a reasonable range do not simply find different communities.

\subsection{Running Time}

\begin{figure}[h!]
\begin{center}
\includegraphics[width=.85\columnwidth]{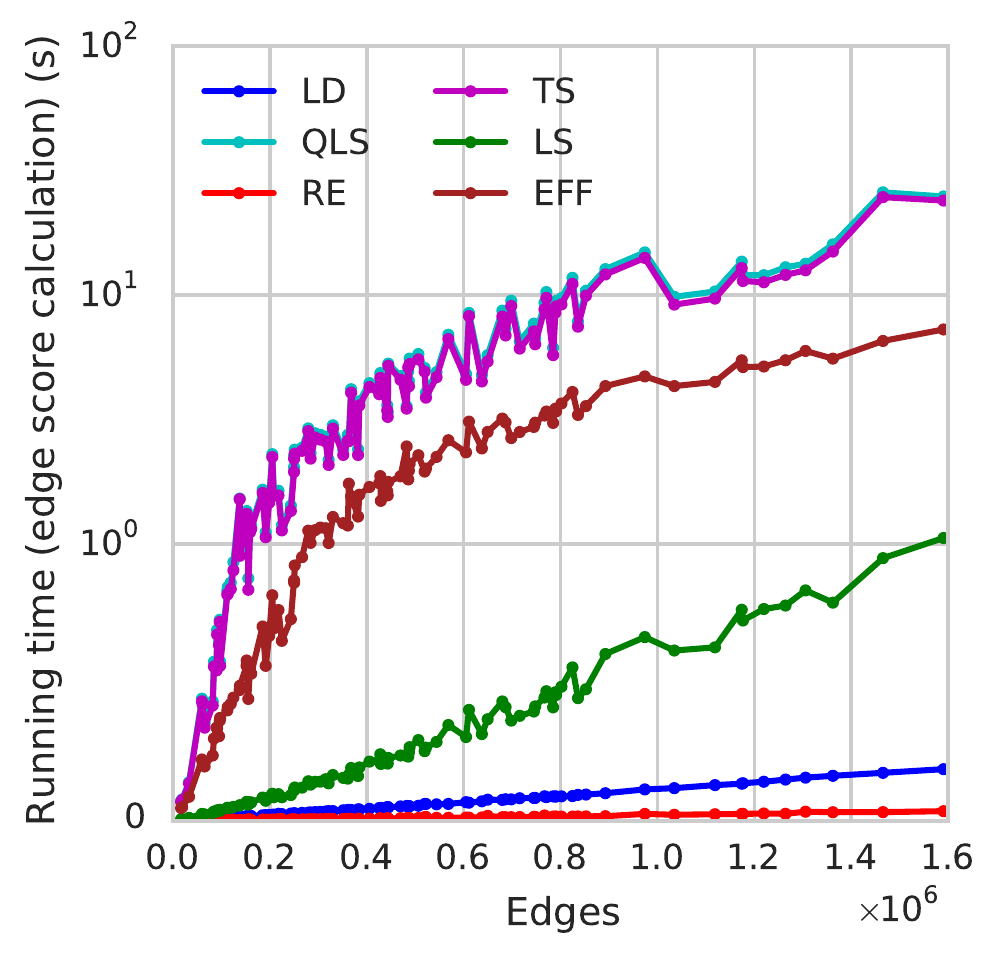}
\caption{Running times of various edge scoring methods on our graph dataset. The y-axis is linearly scaled below 1 and logarithmically above.}
\label{fig:plt_runtimes}
\end{center}
\end{figure}

Measured running times are shown in Fig.~\ref{fig:plt_runtimes}.
Apart from Random Edge sparsification, our Local Degree method is clearly the fastest method and scales linearly with the number of edges, which makes it well suited for large-scale networks in the range of millions to billions of edges.
LS is also fast and could be further accelerated using inexact Jaccard coefficient calculation.
Both Simmelian methods are significantly slower than the other methods, but still efficient enough for the network sizes we consider.
While the time complexity in O-notation of EFF is difficult to assess, it is only slightly faster than the Simmelian methods and not among the fastest methods.

\section{Conclusion}

Our experimental study shows that several methods are capable of preserving a set of relevant properties of social networks when up to 80\% of edges have been removed. 
Random edge deletion performs surprisingly well and retains a wide range of properties, but more targeted methods can perform even better: 
Our novel Local Degree (LD) method seems to be suited for preserving node centralities like degree, PageRank and betweenness.
Also, the clustering, connectedness and the typically small diameter of complex networks are significantly better preserved than through random deletion.
This supports the initial motivation of LD, namely that connections to hubs are highly important for a network's structure.
Only the community structure seems to be discarded by LD, where RE actually performs best of all methods considered.
Furthermore, LD is only slightly more computationally expensive than random edge selection, and therefore applicable to very large networks.
The LS method has been developed to support community detection, and we confirm its suitability for this purpose.
However, network diameter and node centralities tend to get distorted.
Our adaptation of the Forest Fire sampling algorithm to edge scoring fails at preserving node centralities, but is the second best at keeping the network diameter.

We hope that the conceptual framework of edge scoring and filtering as well as our evaluation methods are steps towards a more unified perspective on a variety of related methods that have been proposed in different contexts.
Future developments can be easily carried out within this framework and based on our implementations, which will be available as part of a future release of the open-source network analysis package NetworKit\footnotemark.
\footnotetext{\url{https://networkit.iti.kit.edu/}}

\newpage

\bibliographystyle{IEEEtran}
\bibliography{Bibliography}

\end{document}